\documentclass[aps,prl,twocolumn,amsfonts,showpacs]{revtex4} 
\usepackage{epsfig,amsopn}
\usepackage{graphicx}
\usepackage{amsmath}
\usepackage{natbib}

\def\be{\begin{equation}}
\def\ee{\end{equation}}
\def\bea{\begin{eqnarray}}
\def\eea{\end{eqnarray}}

\def\om{\omega}
\def\nn{\nonumber}

\begin{document}

\title{Driving particle current through narrow channels using classical pump}
\author{Kavita Jain$^1$, Rahul Marathe$^2$, Abhishek Chaudhuri$^2$ and Abhishek Dhar$^2$}
\affiliation{$^1$ Department of Physics of Complex Systems, Weizmann Institute of Science, Rehovot 76100, Israel \\
$^2$ Raman Research Institute, Bangalore 560080, India}
\date{\today}

\begin{abstract}
We study a symmetric exclusion process in which the hopping rates at two
chosen adjacent sites vary periodically in time and have a relative phase 
difference. This mimics a colloidal suspension 
subjected to external time dependent 
modulation of the  local chemical potential. The two special sites act as a classical 
pump by generating an oscillatory current with a nonzero ${\cal DC}$ value 
whose direction depends on the applied phase difference. We
analyze various features in this model through simulations and obtain an 
expression for the $\cal{DC}$ current via a novel perturbative 
treatment.
\end{abstract}
\pacs{05.70.Ln,05.40.-a,05.60.-k,83.50.Ha}
\maketitle 

A number of studies have used the idea of  Thouless adiabatic pumping
\cite{Thouless:1983} to generate   
$\cal{DC}$ current responses to $\cal{AC}$ driving fields in quantum systems.
This mechanism has been used 
experimentally to generate charge \cite{Switkes:1999} and spin 
\cite{Watson:2003} current in systems such as quantum dots and  nanotubes 
and studied theoretically in  various quantum systems without  
\cite{Brouwer:1998}
and with \cite{Aleiner:1998} interactions. 
One may ask if this principle can be used to drive current in 
{\it classical} systems. 
It was shown in \cite{Moskalets:2001} that inelastic scattering can
enhance pumping which suggests that similar effects can be obtained in
purely classical systems. Indeed classical pumping of particles and
heat has been achieved in mesoscopic systems with large inelastic
scattering \cite{Moskalets:2001}, stochastic models
\cite{Astumian:2001,Marathe:2007,Sinitsyn:2007} and seen in experiments 
\cite{Astumian:2003}.
As discussed in \cite{astumian02} these pump models are similar
to  Brownian ratchets \cite{Reimann:2002} where non-interacting 
particles placed in 
spatially asymmetric  potentials that vary periodically in time and
acted upon by noise execute directed motion. However for the model described  
below, particle interactions seem necessary for  pumping.

Here we show how the designing principle of quantum pumps may be applied 
to drive classical particles such as micron-sized charged colloidal particles 
confined in a closed narrow tube. 
Systems of colloidal particles with excluded volume interactions  
diffusing in a narrow channel have been modeled in the recent past 
\cite{Chou:1998,Wei:2000} by one-dimensional 
symmetric exclusion process (SEP) in which hard-core particles attempt to hop 
to  an empty neighbor with equal rate \cite{Liggett:1985}. 
Recently much attention has been given 
to nonequilibrium steady states of driven SEP
\cite{Spohn:1983,Santos:2001}, in which 
particles can enter or leave the system at the boundaries, 
to study large deviation functional \cite{Derrida:2002} and current 
fluctuations \cite{Derrida:2004} in nonequilibrium systems.

In this Letter, we introduce a SEP in which the  hop-out 
rates at some lattice sites are chosen to be time-dependent with a relative 
phase difference between different points. At a coarse-grained level,
this choice models oscillating voltages applied to the colloidal system. 
We find that an oscillatory 
particle current is generated across a bond which interestingly has 
a nonzero $\mathcal{DC}$ value. Unlike driven SEP, our model conserves the 
total number of particles and due to the periodically 
varying  hopping rates, the system reaches in the long time
limit  a periodic time-dependent state.  

Here we will mainly focus on the model with a single pump which we
define as consisting of two adjacent sites where the hopping rates
vary periodically and with a phase difference. The hopping rate at all
other sites is constant. 
Our Monte-Carlo simulations indicate that  the magnitude of the
$\mathcal{DC}$ current $\bar J$ depends sinusoidally on the  
phase difference between the rates at these two sites, and varies
non-monotonically  with 
the frequency of the drive vanishing in both zero and infinite
frequency limits.  
We also obtain an analytical understanding of the system 
by developing a perturbation theory in the strength
of the time-dependent part of the  hopping rates, similar to
\cite{Astumian:1989}. We show that the leading  
order contribution to $\bar J$ is obtained at second order 
in this perturbing parameter, and find an explicit expression for it 
(see Eq.~(\ref{final})) which captures various features seen in the simulations. 
\begin{figure}
\includegraphics[width=3in]{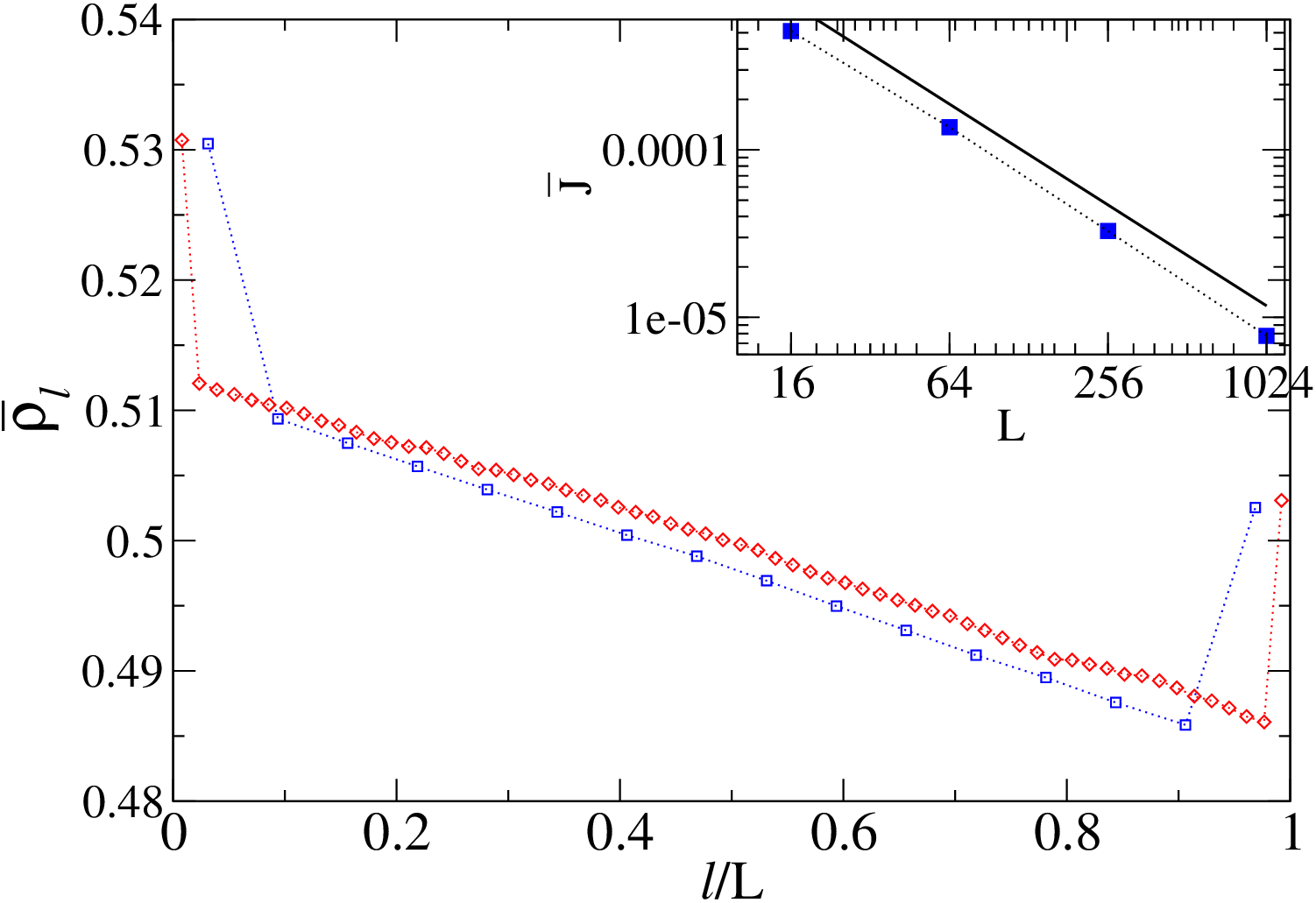}
\caption{(Color online) ${\cal DC}$ density profile $\bar{\rho}_l$ across 
the ring for $f_0=0.3, f_1=0.2$, $\omega=0.2 \pi$ and $\phi=\pi/2$ at 
half filling for two system sizes. Inset: ${\cal DC}$ current 
$\bar{J} \sim 1/L$ as shown by solid line of slope $-1$.}
\label{rhovsi}
\end{figure}

The model is defined on a ring with $L$ sites. A site $l=1,2...,L$ can be 
occupied by $n_l=0$ or $1$ particle, and the system contains 
a total of $N=\rho L$ particles 
where $\rho$ is the density. A 
particle at site $l$ hops to an empty site either on the left or right with 
rate $u_l= f_0$ for all $l \neq 1,L$ and  $u_1=f_0+f_1 \sin(\omega t),~ 
u_L=f_0+f_1 \sin(\omega t+\phi)$.
For $f_1=0$, the above model reduces to the SEP with periodic boundary 
conditions, many of whose properties are known 
exactly \cite{Liggett:1985}. The steady state obeys the equilibrium 
condition of detailed balance so that the current across a bond is zero. 
Due to this, the steady state measure is uniform and the $k$-point correlation 
function 
$C_{l_1, l_2,...,l_k}^{(0)}=\langle n_{l_1} n_{l_2} ... n_{l_k} \rangle={L-k \choose N-k}/{L \choose N}$. As we will describe later, 
this knowledge of the exact 
steady state in the absence of time-dependent term in $u_l$ allows us to 
set up a perturbation expansion in $f_1$ of various observables.

\begin{figure}
\includegraphics[width=3in]{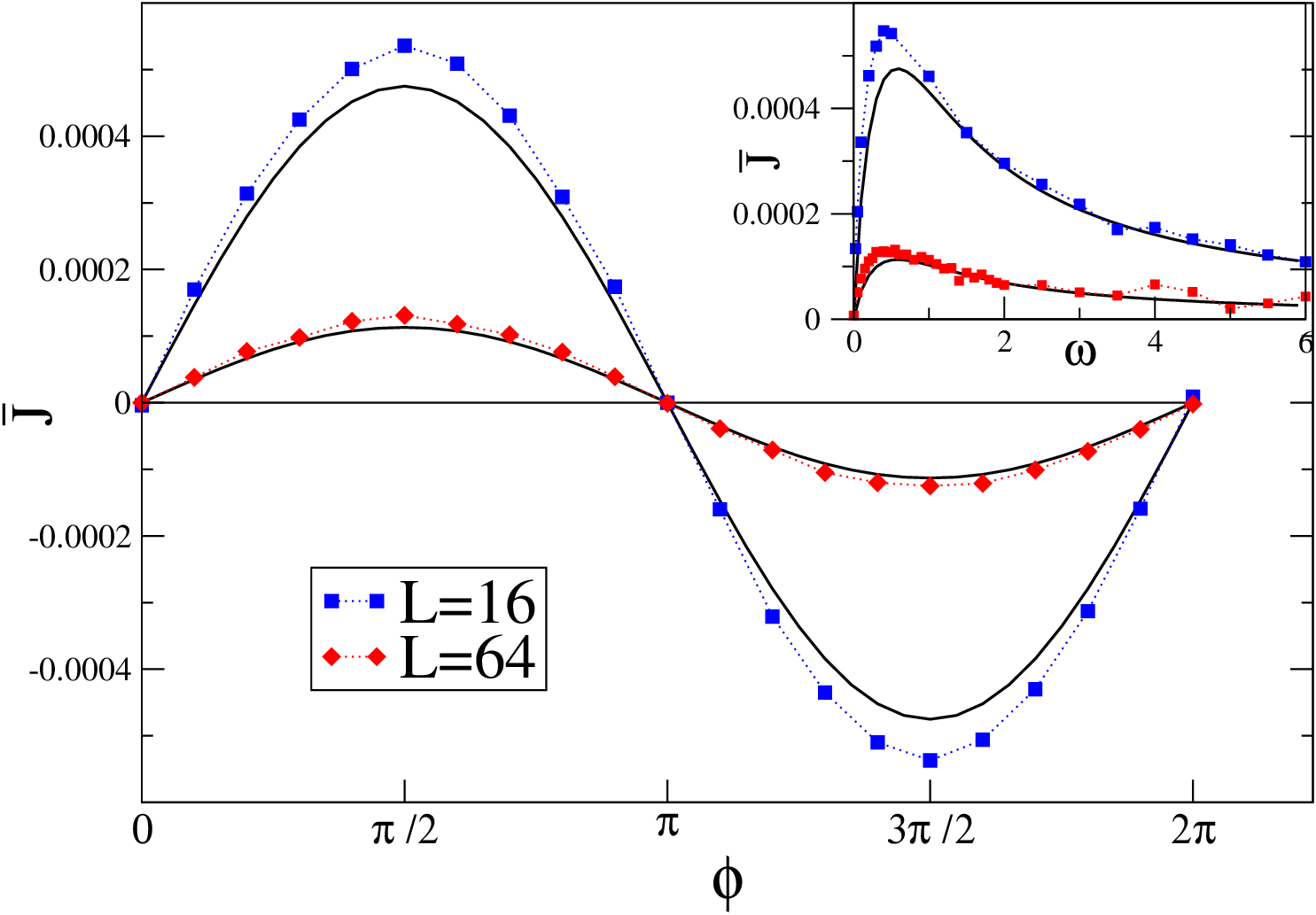}
\caption{(Color online) Plot of 
$\bar{J}$ vs. $\phi$ with other parameters as in Fig.~\ref{rhovsi}.
The solid lines are from the perturbation theory.Inset shows plot of 
$\bar J$ vs. $\om$ }
\label{jvsphi}
\end{figure}
We first discuss the results of our  Monte-Carlo simulations of 
the model defined above for generic values of the driving 
frequency $\omega$ and the phase $\phi$.  
We first look at the 
$\mathcal{DC}$ values of the current and local densities.
In Fig.~\ref{rhovsi}, we have plotted the local particle density  profile for
system sizes $L=16$ and $64$. We find a linear density profile in the bulk
of the system with a discontinuity at the pumping sites. The linear
density profile implies a $\mathcal{DC}$ current in the system. The inset in
the figure shows that this current scales as $1/L$ as in driven SEP 
\cite{Derrida:2002,Derrida:2004}.  
In Fig.~\ref{jvsphi} we plot the current as a function of the
applied phase difference for different system
sizes. The prediction of the perturbation theory (described below) is also
shown for comparison. 
As in the case of quantum pumps, we find a sinusoidal
dependence of the current on the phase $\phi$.  
The inset of Fig.~\ref{jvsphi} shows the dependence of the current on the
driving frequency $\om$ again for $L=16,64$. At $\omega=0$, the system is 
in equilibrium and there is no current \cite{note1}. In the opposite limit 
$\omega \to \infty$, the system does not have time to react to the
rapidly changing fields and we  
may again expect zero current. For any finite $\omega$, the 
${\cal DC}$ current is nonzero and behaves as 
 \be
\bar{J} \sim 
\begin{cases} \omega & \text{~,~$\omega \ll \omega^*$,} \\
1/\omega &\text{~,~ $\omega \gg \om^*$}~, 
\end{cases}
\label{Jomega}
\ee
where $\omega^*$ is the frequency at which the current peaks. 
A similar non-monotonic behavior in current has been seen in experiments 
on ion pumps \cite{Astumian:2003}. 
Since $J \sim \omega$ as  
$\omega \to 0$, this means that a finite number of particles are 
  circulated even in the adiabatic  limit \cite{Astumian:2003}.  
This suggests a perturbation expansion in $\omega$ and will be 
discussed elsewhere. 
We also find that $\omega^*$ scales  linearly with $f_0$, is independent of 
$L$ and hence of the
typical relaxation time $\tau=L^2/f_0$ of the system.  

\begin{figure}
\includegraphics[width=3in]{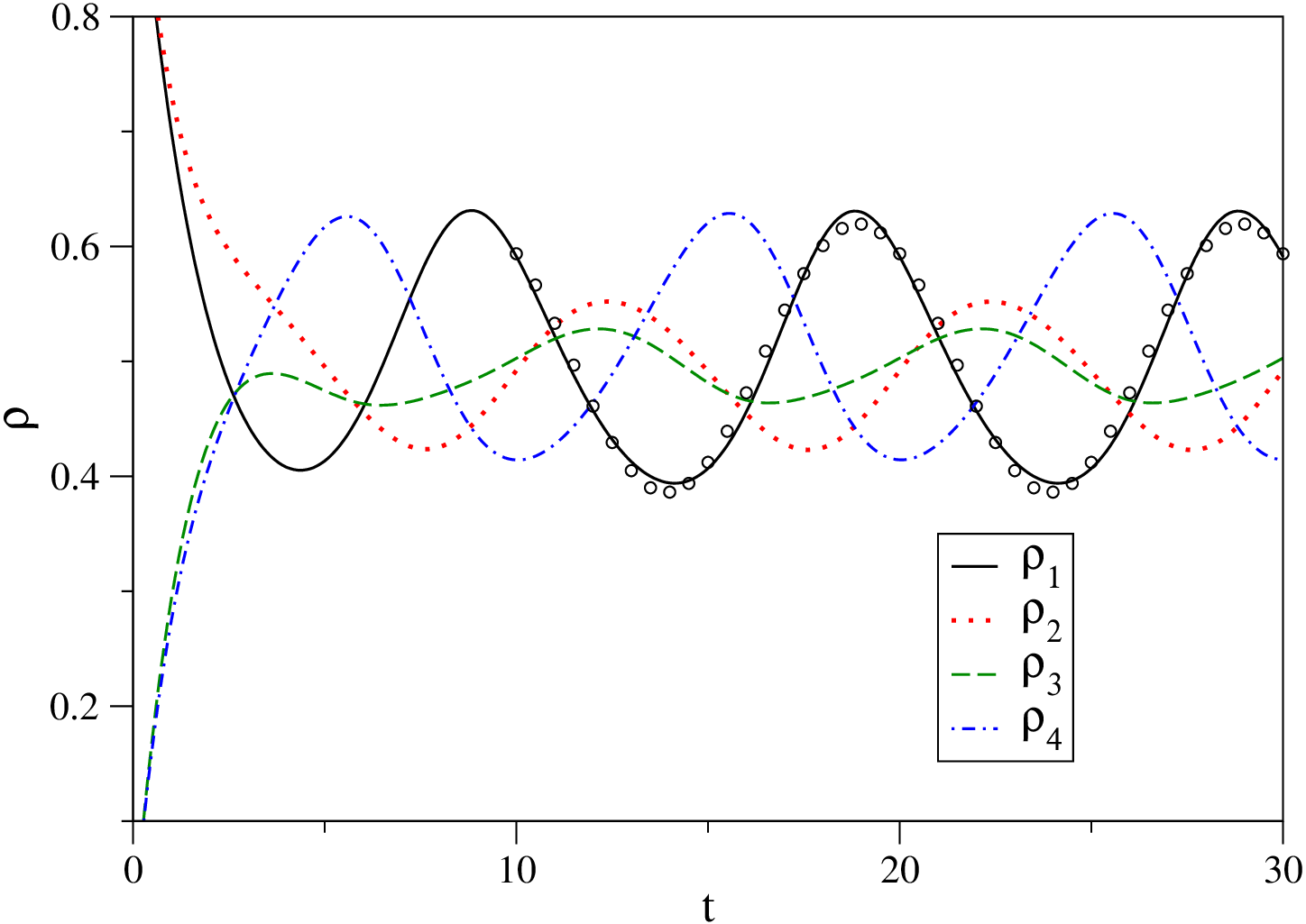}
\caption{(Color online) Plot of time-dependent densities at the four sites of a
  $L=4$ lattice. In the initial configuration, 
 sites $1$ and $2$ have one particle each and other sites are empty. The averages over one 
time period  give: $\bar{\rho}_1 = 0.503493,~   \bar{\rho}_2 =0.498702,~\bar{\rho}_3 =0.497417,~\bar{\rho}_4  = 0.500388$ and $\bar{J}
  =  0.000514$. The points show the curve $\rho+f_1 \rho_1^{(1)}+f_1^2 \rho_1^{(2)}$. [Parameters: $f_0=0.4,~f_1=0.1,~\phi=\pi/2$
  and $\om=0.2\pi$].} 
\label{timeev}
\end{figure}

Till now we have looked at various time averaged quantities. It is
also of interest to look at the time dependences. In
Fig.~\ref{timeev} we show the ensemble averaged local densities
as a function of time obtained by an exact numerical integration of
the $L=4,~N=2$ master equation. We find that the local densities oscillate 
about their respective ${\cal DC}$ value ( different from $\rho=0.5$)
at large times,  
and a nonzero ${\cal DC}$ current circulates around the ring. Although the local ${\cal DC}$ density is close to 
$\rho$, the amplitude of the $\mathcal{AC}$ part is quite large. We also show a
comparison of the exact density $\rho_l(t)$ 
with the approximate density $\rho+f_1 \rho_l^{(1)}+f_1^2 \rho_l^{(2)}$
obtained from the perturbation theory 
(see Eqs.~(\ref{rho1_t}) and (\ref{rho2_t})). We find that the ${\cal AC}$ 
density contains the driving frequency and also higher harmonics.

We now proceed to describe a 
perturbation theory within which we will calculate an expression for the 
$\mathcal{DC}$ current $\bar J$ in the bulk of the system. Let us denote the 
site hopping rates by
\be
u_l = f_0 +f_1 v_l =f_0+f_1 (\nu_l e^{i \omega t}+\nu_l^* e^{-i \omega t})~, 
\label{rates_t}
\ee
where the site-dependent complex amplitude 
$\nu_1=-i/2, \nu_L=-ie^{i\phi}/2$  and all other $\nu_l$'s are zero. 
We consider perturbative expansions of various quantities of interest with 
$f_1$ as the perturbation parameter about the homogeneous steady state 
corresponding to $f_1=0$. Thus 
we have
\bea
\rho_l(t) &=& \langle n_l (t) \rangle = \rho+ \sum_{k=1}^{\infty} f_1^k 
\rho_l^{(k)} (t)  \nn \\
C_{l,m}(t) &=& \langle n_{l}(t) n_{m} (t)\rangle = C_{l,m}^{(0)}+
\sum_{k=1}^{\infty} f_1^k  C_{l,m}^{(k)} (t)~,   \label{pert}
\eea
and  similar expressions for higher correlations. The current in a bond 
connecting sites $l$ and $l+1$ is given by 
$J_{l,l+1}(t) = u_l ( \rho_l-C_{l,l+1}) - u_{l+1} (\rho_{l+1}-C_{l,l+1})$ which can 
also be expanded in a perturbation series using 
Eq.~(\ref{pert}). Averaging $J_{l,l+1}(t)$ over a 
time period for $\l \neq L-1, 1, L$, we obtain constant ${\mathcal DC}$ current in the bulk \cite{note2}, 
\be
\bar{J} =f_0 \sum_{k=1}^{\infty} 
f_1^k \;(\bar{\rho}_l^{(k)}-\bar{\rho}_{l+1}^{(k)})
\label{current-pert}
\ee
where $\bar{\rho}_l^{(k)}$ is the time-averaged density at the $k^{\rm th}$ 
order. 
To obtain $\bar{J}$ to leading order in $f_1$, we start with the exact equation 
of motion for the density $\rho_l$ \cite{Schutz:2000} 
\bea
&&\frac{\partial \rho_l}{\partial t}+2 u_l \rho_l- u_{l-1} \rho_{l-1}-u_{l+1} \rho_{l+1} \nonumber \\
 &=&  u_l (C_{l-1,l}+C_{l,l+1}) -u_{l+1} C_{l,l+1}-u_{l-1}C_{l-1,l}   ~.   
 \label{exact}
\eea
Then the density $\rho_l^{(k)}$ at the $k^{\rm th}$ order evolves according to 
\bea
&&\hspace{-0.5cm}\frac{\partial \rho_l^{(k)} }{\partial t}-{f_0} 
\Delta_l  \rho_{l}^{(k)}+ 2 v_l \rho_l^{(k-1)}- v_{l-1} \rho_{l-1}^{(k-1)}- v_{l+1} \rho_{l+1}^{(k-1)}\nn \\
&=& v_l (C_{l-1,l}^{(k-1)}+C_{l,l+1}^{(k-1)})- 
v_{l-1} C_{l-1,l}^{(k-1)}- v_{l+1} C_{l,l+1}^{(k-1)}  ~,    \label{rhok}
\eea
where $\Delta_l g_l=g_{l+1}-2g_l+g_{l-1}$ defines the discrete
Laplacian operator. Thus, the density at the $k^{\rm th}$ order is 
obtainable in terms of the density and the 
two-point correlation function at $(k-1)^{\rm th}$ order. 

At first order, the above equation simplifies to give 
\be
\frac{\partial \rho_l ^{(1)} }{\partial t}-{f_0} \Delta_l \rho_l^{(1)} 
=r_0 \Delta_l v_l~, \label{rho1}
\ee
where $r_0=\rho-C_{l_1,l_2}^{(0)}$. In the steady state, integrating
both sides of Eq.~(\ref{rho1}) over a time period,   
and using $\rho_l ^{(1)}=\rho_{l+L} ^{(1)}$ and density conservation, we find 
$\bar{\rho}_l ^{(1)}=0$. Hence at ${\cal{O}}(f_1)$, no $\cal{DC}$ current 
is generated. This is consistent with our expectation from linear response
 theory, namely that an $\cal{AC}$ field gives an $\cal{AC}$ response.   
Thus, 
we need to consider the next order term $\rho_l^{(2)}(t)$ in the 
perturbation expansion of the density to obtain current. As 
Eq.~(\ref{rhok}) for $k=2$ involves density at first order, we first solve for 
$\rho_l ^{(1)}(t)$ in Eq.~(\ref{rho1}). The solution of this inhomogeneous 
linear equation is a sum of a homogeneous part which depends on
initial conditions, and a particular integral. At long times, the
homogeneous part 
can be shown to vanish (using the condition $\sum_l
\rho_l^{(1)}(0)=0$) while the particular integral has the asymptotic form
\be
\rho_l ^{(1)}(t) =  A_l^{(1)} e^{i \omega t} + A_l^{*(1)} e^{-i \omega
  t} \label{rho1_t}
\ee
with the amplitudes given by 
\bea
A_l^{(1)}&=&\alpha_+z_+^l+\alpha_- z_-^l ~~,~~l=2, 3..., L-1 \nn \\
A_1^{(1)}&=& \alpha_+z_++ \alpha_- z_- -\frac{r_0 \nu_1}{f_0}\nn \\
A_L^{(1)}&=& \alpha_+z_+^L + \alpha_- z_-^L -\frac{r_0 \nu_L}{f_0}\nn~
\eea
where $z_+=y/2+[(y/2)^2-1]^{1/2},~z_-=1/z_+,~y=2+i \om/f_0$ 
and $\alpha_\pm=(-i r_0 \omega/f_0^2)(\nu_1 z_\mp +
\nu_L)/[(z_{\pm}^L-1)(z_\mp-z_{\pm})]$. The amplitude of the density
oscillations decay exponentially with distance from the pumping sites.

We also need  the two-point correlation function $C_{l,m}^{(1)}$ 
at ${\cal O}(f_1)$ appearing in Eq.~(\ref{rhok}) for $k=2$. 
Due to hard-core interaction, the correlation function 
$C_{l,m}$ for nearest neighbors $l$ and $m$ 
obeys a different equation of motion when $|l-m| \neq 1$. Taking 
care of this complication, it is straightforward to write down the 
corresponding 
evolution equations \cite{Schutz:2000} and carry out the perturbative 
expansion. 
At ${\cal O}(f_1)$ in perturbation theory, we then get (for $ 1\leq l
< L,~ l+1 <m \leq L$) 
\bea
&&\hspace{-0.5cm}\frac{\partial C_{l,m}^{(1)}}{\partial t}- f_0
(\Delta_l +\Delta_m) C_{l,m}^{(1)} = k_0(\Delta_l v_l + \Delta_m v_m)~,\nn
 \\ 
&&\hspace{-0.5cm}\frac{\partial C_{l,l+ 1}^{(1)}}{\partial t}+ f_0 \left( 2 C_{l,l +
    1}^{(1)} -C_{l-1,l+1}^{(1)}-C_{l,l + 2}^{(1)} \right) \nn \\
&&~~~~~~~~~~~~~~~~~~~~~~= k_0(v_{l-1}+v_{l+2}-v_l-v_{l+1})  
\label{corr-all}
\eea
where $k_0=C_{l_1,l_2}^{(0)}-C_{l_1,l_2,l_3}^{(0)}$.
The computation of even the homogenous solution of the above set of equations 
is in general a nontrivial task because of the form of the equations
involving nearest neighbor indices and requires a Bethe ansatz or
dynamic product ansatz \cite{Schutz:2000,Santos:2001}. However the
long time solution can still be found exactly and is given by:
\be
C_{l,m}^{(1)}(t)=\frac{k_0}{r_0} [\rho_l^{(1)}(t)+\rho_m^{(1)}(t)]= A_{l,m}^{(1)} e^{i \omega t} + A_{l,m}^{*(1)} e^{-i \omega t}~,
\label{corr-soln}
\ee
where $A_{l,m}^{(1)}=(k_0/r_0)(A_l^{(1)}+A_m^{(1)})$. 
It is easily verified that this satisfies Eq.~(\ref{corr-all}) for all 
$l, m$.
Interestingly, if one were to assume that the two-point function decouples,  
then a perturbation expansion to order $f_1$ will also yield the preceding 
equation in the thermodynamic limit. 
To determine whether the system indeed has a product measure 
requires 
a  more detailed analysis of the higher order terms in the perturbation series 
and higher correlations, and will be discussed elsewhere. 

We are now in a position to find the density profile $\rho_l^{(2)}(t)$. Using 
Eqs.~(\ref{rho1_t}) and (\ref{corr-soln}) in Eq.~(\ref{rhok}) for $k=2$, we 
get  
\be
\rho_l ^{(2)}(t) = \bar{\rho}_l^{(2)}+ A_l^{(2)} e^{i 2 \omega t} + A_l^{*(2)} e^{-i 2 \omega t} \label{rho2_t}
\ee
where the ${\cal DC}$ part $\bar{\rho}_l^{(2)}$ obeys the equation 
\bea
&&f_0 \Delta_l \bar{\rho}_l^{(2)}=2~{\rm Re}[2 \nu_l^* A_l^{(1)}- \nu_{l-1}^* A_{l-1}^{(1)}- \nu_{l+1}^* A_{l+1}^{(1)} \nonumber \\
&&-\nu_l^* (A_{l-1,l}^{(1)}+A_{l,l+1}^{(1)})+\nu_{l-1}^* A_{l-1,l}^{(1)}+ \nu_{l+1}^* A_{l,l+1}^{(1)}]~.
\eea
Noting that the right hand side of the above equation is zero for most $l$, we
get the following solution
\bea
\bar{\rho}_l^{(2)}&=&{s}l+h  ~,~l=2,...,L-1 \nn\\
\bar{\rho}_1^{(2)}&=&{s}+h+\frac{2}{f_0}~{\rm Re}[\nu_1^* (A_{1,2}^{(1)}-A_1^{(1)})] \nn \\
\bar{\rho}_L^{(2)}&=&{s}L+h+\frac{2}{f_0}~{\rm Re}[\nu_L^* (A_{L-1,L}^{(1)}-A_L^{(1)})]~,
\eea
where the slope $s$ of the linear density profile is given by
\be
s=\frac{2}{L f_0} ~{\rm Re}[\nu_1^* (A_{1,2}^{(1)}-A_{1,L}^{(1)})+
\nu_L^* (A_{1,L}^{(1)}-A_{L-1,L}^{(1)}) ]~, \label{eqslope}
\ee
and the intercept $h$ can be found using the particle conservation 
condition $\sum_l \rho_l^{(2)}=0$. 

Our final result for the leading order contribution to the $\cal{DC}$ 
current in Eq.~(\ref{current-pert}) is $\bar{J}=-f_0 f_1^2 s$ and,
using the form of $A_{l,m}^{(1)}$, can be written as
\be
\bar J= -\left( \frac{f_1}{f_0} \right)^2 \frac{k_0 \omega \sin \phi}{2 L} ~{\rm Re}\left( \frac{z_+-z_+^{L-1}}{1-z_+^L} \right) + {\cal O}(f_1^3)~. \label{final}
\ee
We observe that the $1/L$ dependence of the 
$\mathcal{DC}$ current and its 
sinusoidal variation with the phase are captured by the perturbation theory 
at ${\cal O}(f_1^2)$. 
One can also check from the above expression that the frequency
dependence of current in Eq.~(\ref{Jomega}) is obtained with the
typical frequency $\omega^* \approx 2 f_0,~ L \gg 1$. 
As shown in Fig.~(\ref{jvsphi}) the above
perturbative solution is already in a very good quantitative agreement with the
simulation results.

To summarise, we have studied a stochastic model of hard-core
particles on a ring in which  periodically varying hopping rates 
can induce a $\cal{DC}$ current. 
A variation of our model defined on an open chain connected to 
particle reservoirs also shows nonzero $\cal{DC}$ current. 
Simulations and a novel perturbative
analysis presented here
show that many of the qualitative features are similar to
that seen in quantum pumps.  However unlike in quantum pumps, interactions
play an essential role in our model. 
For non-interacting particles, noting that the bond current 
$J_{l,l+1}=u_l \rho_l -u_{l+1} \rho_{l+1}$ summed over all bonds is zero 
and using the uniformity of 
$\bar J$, it follows that no ${\cal DC}$ current is generated in this case. 
Our model can be generalised to include several pumps and pumps consisting of 
several sites, and these features could increase pumping efficiency.  

K.J. thanks A. R\'akos for useful discussions,  Raman Research
Institute for kind hospitality and Israel Science Foundation for
financial support. A.D. thanks J.L.Lebowitz and E.R.Speer for useful
discussions.  



\begin{thebibliography}{8}
\bibitem{note2} That ${\cal DC}$ current $\bar J$ is same in all bonds follows on using that the density is conserved over a period in Eq.~(\ref{exact}).
\end{thebibliography}
\end{document}